\documentclass[twocolumn, prl, superscriptaddress]{revtex4-2}

\usepackage{graphicx}% Include figure files
\usepackage{dcolumn}% Align table columns on decimal point
\usepackage{bm}% bold math
\usepackage{anyfontsize} %
\usepackage{colortbl}   % For coloring table lines
\usepackage{xcolor}     % For color definitions
\usepackage{steinmetz}

\usepackage{amsmath,graphicx,bbold}
\usepackage{dcolumn}   % needed for some tables
\usepackage{bm}        % for math
\usepackage{color}
\usepackage{physics}
\usepackage{ulem}
\usepackage{dblfloatfix}
\usepackage{float}
\usepackage{comment}
\usepackage{lineno}

\newcommand{\e}{\mathrm{e}}

\newcommand{\bpm}{\begin{pmatrix}}
\newcommand{\epm}{\end{pmatrix}}

\newcommand{\be}{\begin{equation}}
\newcommand{\ee}{\end{equation}}
\newcommand{\bea}{\begin{eqnarray}}
\newcommand{\eea}{\end{eqnarray}}

\graphicspath{{./images/}}

\begin{document}

\title{Measuring Geometric Phase based on Indefinite Causal Order in a Sagnac Interferometer}

\author{Lucas Marques Fagundes}
\thanks{These authors contributed equally to this work.}
\affiliation{Departamento de F\'{i}sica, Universidade Federal de Santa Catarina, CEP 88040-900, Florian\'{o}plis, SC, Brazil}

\author{Finlay Campbell}%
\thanks{These authors contributed equally to this work.}
\affiliation{School of Physics and Astronomy, University of Glasgow, Kelvin Building, Glasgow G12 8QQ, UK}

\author{Richard Aguiar Maduro}%
\affiliation{School of Physics and Astronomy, University of Glasgow, Kelvin Building, Glasgow G12 8QQ, UK}

\author{Renn\'e Medeiros de Ara\' ujo}%
\affiliation{Departamento de F\'{i}sica, Universidade Federal de Santa Catarina, CEP 88040-900, Florian\'{o}plis, SC, Brazil}

\author{Stephen M. Barnett}%
\affiliation{School of Physics and Astronomy, University of Glasgow, Kelvin Building, Glasgow G12 8QQ, UK}

\author{Sarah Croke}%
\affiliation{School of Physics and Astronomy, University of Glasgow, Kelvin Building, Glasgow G12 8QQ, UK}

\author{Sonja Franke-Arnold}%
\affiliation{School of Physics and Astronomy, University of Glasgow, Kelvin Building, Glasgow G12 8QQ, UK}

\date{\today}% It is always \today, today,

\begin{abstract}

Sagnac interferometers are important tools for precision measurements. Cancellation of common-path noise ensures a high degree of stability against external perturbations, while their sensitivity to rotations provides the basis of laser gyroscopes. Here we highlight another, less explored feature of Sagnac interferometers: the fact that optical elements are traversed in reverse order for clockwise and counterclockwise propagating light components. This provides the opportunity to explore a classical realization of indefinite causal order, where the order of events within a sequence is not fixed. We explore this concept to identify the geometric phase associated with two non-commuting polarization operations with a single measurement. This idea may have applications for the rapid determination of polarization manipulations or optical activity within chiral media, but foremost it provides a geometric illustration of indefinite causal order.

\end{abstract}

\maketitle
% \def\thefootnote{*}\footnotetext{These authors contributed equally to this work}
% \def\thefootnote{\arabic{footnote}}text text text\footnote{normal footnote}
%\tableofcontents
% \def\thefootnote{$\dag$}\footnotetext{These authors contributed equally to this work}

\section{Introduction}
% \cite{test_min_working_example}
The geometric phase is a fundamental property of wave evolution, arising from the geometry of the trajectory traced by the system's state in its parameter space, rather than from the dynamical details of the evolution. Since the seminal work of Pancharatnam \cite{Pancharatnam1956} and later Berry \cite{BerryPhase1984,BerryPhase1987}, geometric phases have become a central concept in optics and quantum physics, providing insights into wave interference \cite{bhandari1988observation}, polarization manipulations \cite{rao2026evolution,rodrigues2024resonance} including polarization control in optical fibers \cite{lefevre1983fiber, yang2023geometric}, topological phenomena \cite{bliokh2019geometric}, and quantum state evolution \cite{Cohen2019GeometricPhaseReview}. Geometric phase devices can be used for a wide range of light manipulation including beam shaping, generating orbital angular momentum beams and holography \cite{ Cohen2019GeometricPhaseReview, GeometricPhaseReviewJisha2021, Faraz2025GeometricPhaseReview}.  In optical polarization geometric phases can result in the 
rotation or twisting of the Poincar\'{e} vector \cite{Pancharatnam1956,BerryPhase1987}.  It is 
this feature that we explore here within the framework of indefinite causal order.

Geometric phases are attractive for applications in precision metrology \cite{Zhou2025Metrology}, as they tend to be less sensitive to fluctuations that would affect the dynamical phase such as intensity noise or timing jitter. Furthermore, geometric phases may amplify small parameter changes as their path-dependence can be non-linear, and often can be defined in a gauge-invariant way, {\it i.e.}~independent of initial and dynamic phases. This motivates the development of techniques capable of measuring geometric phases with high sensitivity. 
Experimentally, these benefits are best realized in Sagnac interferometers.

Some of the authors have recently linked geometric phases encountered by a sequence of unitary transformations of polarization states to the concept of indefinite causal order (ICO)\cite{SagnacICO}. ICO describes situations, where the order of quantum events or operations is not fixed but exists in a superposition of different sequences. Sagnac interferometers provide the natural tool to realize geometric phase measurements based on ICO, as operations placed within the interferometer arms are encountered in opposite order by the co- and counterpropagating beams.

ICO was originally introduced in the context of general relativity \cite{Hardy2009} and later formalized within quantum information theory by Chiribella \textit{et al.} \cite{chiribella2013quantum}. Since then, ICO protocols have attracted considerable attention because of their potential advantages in quantum information processing, including reductions in quantum gate-query complexity \cite{escandon2023practical,zhao2025unambiguous}, time-reversal universal computation \cite{schiansky2023demonstration,simonov2026universal}, efficient phase estimation \cite{taddei2021computational}, and other applications in quantum metrology \cite{ICOMetrology2020Zhao,ICOMetrology2021Chapeau-Blondeau}. Several of these advantages have already been demonstrated experimentally \cite{procopio2015experimental,rubino2017experimental,goswami2018indefiniteExperiment,wei2019experimental,ICOexperimental2023Yin,stromberg2023experimental,cao2023Experimental,rozema2024experimental,stromberg2024experimental}.

Here we present an experimental realization of the ICO protocol introduced in \cite{SagnacICO} to measure the geometric phase of a sequence of polarization operations.
While ICO is generally investigated in the context of quantum superpositions of different sequences of operations, we show that measurement advantages also persist in a classical context. 

Our protocol seeks to measure the geometric phase 
associated with two unknown unitary polarization transformations $\widehat{U}$ and $\widehat{V}$, corresponding to different paths on the Poincar\'e sphere, and experimentally implemented by different waveplates. Unless $\widehat{U}$ and $\widehat{V}$ commute, the acquired optical phase will depend on the order in which the transformations are encountered. For an input state, where the optical paths $\widehat{U}\widehat{V}$ and $\widehat{V}\widehat{U}$ reach the same point on the Poincar\'e sphere, their global phase differs by $\Omega/2$, where $\Omega$ is the solid angle of the geodesic path \cite{Pancharatnam1956,BerryPhase1987,GeodesicRule} enclosed by $\widehat{U}^\dagger \widehat{V}^\dagger \widehat{U}\widehat{V}$.

If we have access to $\widehat{U}$ and $\widehat{V}$ separately, then a conventional strategy to determine the geometric phase would rely on identifying the 16 elements of each of the two Mueller matrices. Experimentally, determining each matrix would require at least 16 measurements, but is typically and more conveniently realized with 36 Stokes tomography measurements, testing input states with horizontal, vertical, diagonal, antidiagonal, right and left handed polarization and performing projection measurements in the analogous state basis.

Using the ICO protocol we can determine the accumulated geometric phase in a one-shot-measurement from detecting orthogonal polarizations in two output ports of a Sagnac interferometer.
%without requiring a complete characterization of either Jones matrix. The measurement is based on the output of a Sagnac interferometer, which exploits the coherent superposition of the two possible operation orders to determine the geometric phase.
%exploits the fact that the order of $\widehat{U}$ and $\widehat{V}$ is reversed for the two counter-propagating paths, yielding an expression for the geometric phase. This approach enables the measurement of the geometric phase without prior knowledge of the two waveplates and without requiring preparation of a polarization state that is an eigenstate of $\widehat{U}^\dagger \widehat{V}^\dagger \widehat{U}\widehat{V}$. When the Jones matrices of the waveplates are known, the geometric phase can instead be obtained directly from the eigenvalues of $\widehat{U}^\dagger \widehat{V}^\dagger \widehat{U}\widehat{V}$. Here, we present experimental results for this setup and compare them with the expected geometric phases for known waveplates.
In the following we describe our measurement strategy in detail, present experimental results for several waveplate sequences and compare our measurements with the expected geometric phases.

%There are proposed examples of using an indefinite causal order to gain an advantage in metrology \cite{ICOMetrology2020Zhao,ICOMetrology2021Chapeau-Blondeau} and quantum communication \cite{ICOadvantage2012Chiribella,ICOAdvantage2015Feix,ICOAdvantage2016Guerin,ICOAdvantage2018Ebler}. Several of these systems have also been demonstrated experimentally \cite{procopio2015experimental,rubino2017experimental,goswami2018indefiniteExperiment,wei2019experimental,ICOexperimental2023Yin,stromberg2023experimental,cao2023Experimental,rozema2024experimental,stromberg2024experimental}. The experiment presented here also proposes a measurement that takes advantage of probing a system in multiple orders of operations. 
\section{Indefinite causal order and geometric phase in a Sagnac interferometer}

\begin{figure}
    \centering
    \includegraphics[width=\linewidth]{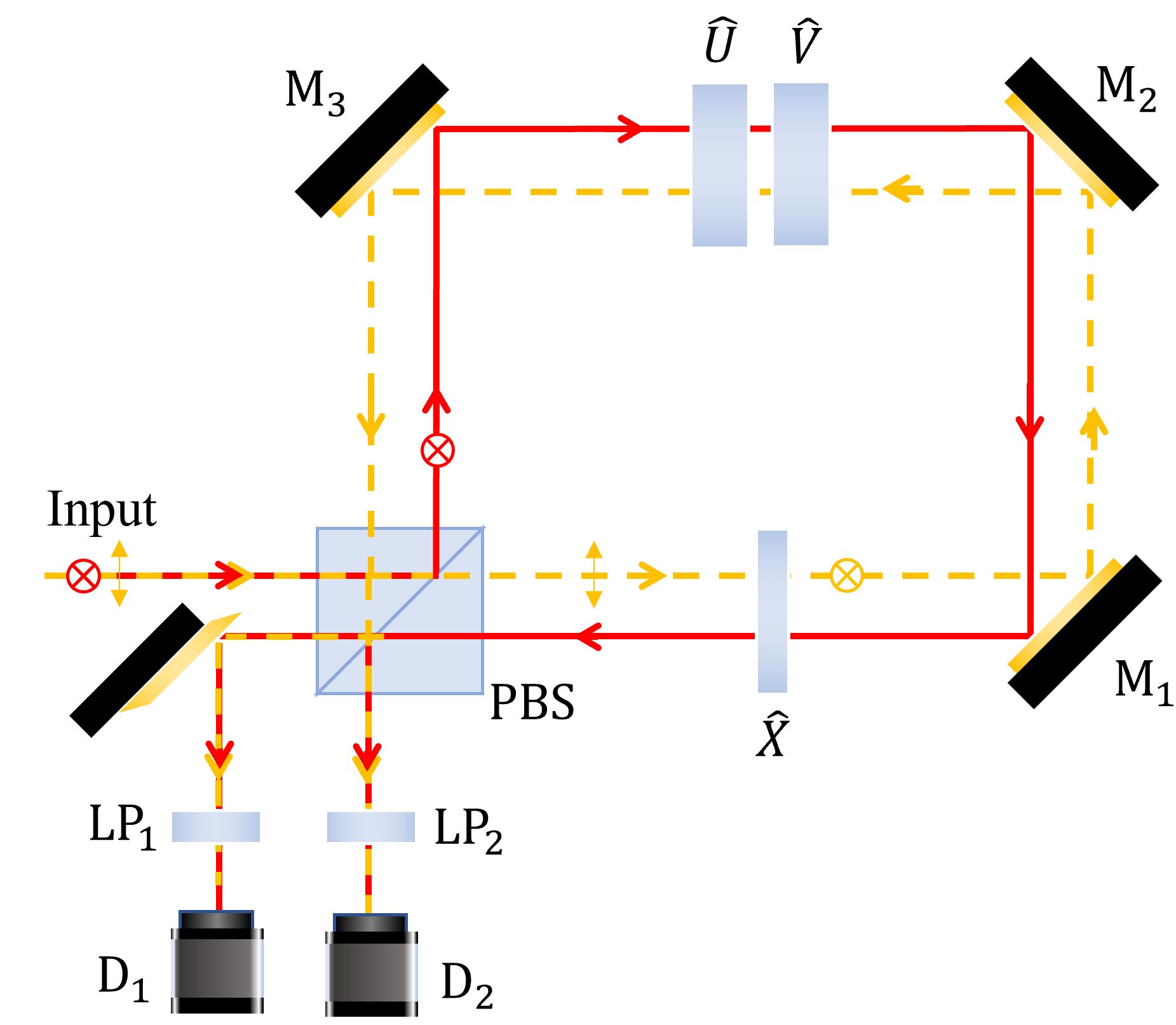}
    \caption{Experimental ICO setup for measuring geometric phases. A diagonally polarized laser beam at 633 nm is sent into a Sagnac interferometer, with the vertical polarization component $|v\rangle$ (shown in red) propagating clockwise and the horizontal component $|h\rangle$ (dashed yellow) anticlockwise. The interferometer comprises two unitary operations of interest for the measurement of geometric phases, realized by waveplates, denoted by $\widehat{U}$ and $\widehat{V}$. A third (half) waveplate, denoted by $\widehat{X}$, is placed in the transmitted arm to swap $\ket{h}$ and $\ket{v}$, ensuring that the polarization state incident on the waveplates is identical for both propagation paths. At the two output ports, the states are analyzed using  linear polarizers and recorded with a CMOS camera.}
    \label{fig:experimental_setup}
\end{figure}

A Sagnac interferometer allows the generation of a coherent superposition of light traversing the interferometer path clockwise and counterclockwise. A sequence of optical elements placed within the interferometer is encountered by these two light components in opposite orders, and moreover every individual element is traversed in the opposite direction. Sagnac interferometers can be constructed using either a non-polarizing or a polarizing beam splitter as their input and exit port. In our experiment, shown in Fig.~\ref{fig:experimental_setup}, we have chosen a polarizing beam splitter (PBS), which means that the clockwise and counterclockwise components are distinguishable in their linear polarization. %This is one of the cases presented by some of the authors in \cite{SagnacICO}. 
We have implemented a displaced Sagnac interferometer with a small separation between clockwise and anti-clockwise interferometer arms, so that we can access both output ports. This means that we rely on the fact that optical transformations of custom waveplates are homogeneous across the optical element. 

For convenience, we represent the polarization and path degrees of freedom in Dirac notation. While the experiment was performed with laser beams rather than single photons, classical and quantum operation is identical for interferometers using only one input port \cite{Barnett_2022_ClassicalQuantumConnection}.

Diagonally polarized input light is incident on the PBS,
so that horizontally polarized light propagates  in the anticlockwise direction (labeled as the state $|A\rangle$ and indicated in red in Fig.~\ref{fig:experimental_setup}) and vertically polarized light in the clockwise direction ($|C\rangle$, shown in yellow). After the first interaction with the PBS, the optical field is therefore described by the state
\begin{equation}
\frac{1}{\sqrt{2}}(|A\rangle|h\rangle + |C\rangle|v\rangle)\,,
\label{eq:0h-1v}
\end{equation}
where $|h\rangle$ and $|v\rangle$ denote horizontal and vertical polarization, respectively. We choose coordinate systems in each arm such that the positive z axis is always along the direction of propagation. To keep a right handed coordinate system we choose the vertical axis to be fixed and the horizontal axis flips sign depending on propagation direction. A waveplate denoted in the clockwise direction as $\widehat{U}$, will therefore in the anticlockwise direction be represented by $-\widehat{Z}\widehat{U}(-\widehat{Z})=\widehat{Z}\widehat{U}\widehat{Z}$ to keep the coordinates consistent. Here $\widehat{Z}$ is the Pauli-$Z$ matrix where $\widehat{Z}|h\rangle = +|h\rangle$ and $\widehat{Z}|v\rangle =-|v\rangle$. 

Our setup contains the two unknown waveplates $\widehat{U}$ and $\widehat{V}$, whose geometric phase $\widehat{U}^\dagger \widehat{V}^\dagger \widehat{U}\widehat{V}$ we wish to determine, along with a third known waveplate $\widehat{X}$, chosen to perform the Pauli-$X$ transformation such that $\widehat{X}|h\rangle = |v\rangle$ and $\widehat{X}|v\rangle = |h\rangle$. This is required as the PBS correlates the path degree of freedom with polarization degree of freedom. The additional Pauli transformation means, that the combination of unknown waveplates is tested by light in the same polarization for both paths. Note that if $\widehat{X}$ is traversed in anticlockwise direction the same mathematical operation discussed above needs to be applied.

After the action of all of the waveplates inside the interferometer, the state (\ref{eq:0h-1v}) becomes
\begin{equation}\begin{split}
    |A\rangle \left(\widehat{U}\widehat{V}\widehat{X} |h\rangle\right)+|C\rangle \left((\widehat{Z}\widehat{X}\widehat{Z})(\widehat{Z}\widehat{V}\widehat{Z})(\widehat{Z}\widehat{U}\widehat{Z})|v\rangle \right) \\
    =|A\rangle \left(\widehat{U}\widehat{V} |v \rangle \right)-|C\rangle \left(\widehat{Z}\widehat{X}\widehat{V}\widehat{U} |v\rangle \right)\,.
\end{split}\end{equation}
The light then leaves the Sagnac interferometer through the PBS. We label the output port combining the transmitted clockwise path and the reflected anticlockwise path as $|1\rangle$ and the other output port as $|2\rangle$. This second pass through the PBS results in an output state given by
\begin{align}
|\Phi\rangle =&\frac{1}{\sqrt{2}}|1\rangle \Bigg(|v\rangle \langle v| \widehat{U}\widehat{V}|v\rangle -|h\rangle 
\underbrace{\langle h|\widehat{Z}\widehat{X}}_{\langle v|}\widehat{V}\widehat{U}|v\rangle \Bigg)
  \nonumber \\
&+
\frac{1}{\sqrt{2}}|2\rangle \Bigg(|h\rangle \langle h| \widehat{U}\widehat{V}|v\rangle -|v\rangle 
\underbrace{\langle v|\widehat{Z}\widehat{X}}_{-\langle h|}\widehat{V}\widehat{U}|v\rangle \Bigg).
\end{align}

The output at each port is then measured in the diagonal basis, $|a\rangle = (|h\rangle  -|v\rangle)/\sqrt{2}$ and $|d\rangle = (|h\rangle + |v\rangle)/\sqrt{2}$. 
%so that path cannot be distinguished by this measurement. This requires that the beams arising from clockwise and anticlockwise propagation are perfectly superimposed at the detector.
In this basis, the output state is, finally,
\begin{align}
    |\Phi\rangle =&\ \frac{1}{2}\,|1\rangle\left( -|a\rangle \langle v| \widehat{U}\widehat{V}+\widehat{V}\widehat{U}|v\rangle +|d\rangle \langle v|\widehat{U}\widehat{V}-\widehat{V}\widehat{U}|v\rangle \right)  \nonumber \\
    &\ 
    +\frac{1}{2}\,|2\rangle\left(|a\rangle \langle h| \widehat{U}\widehat{V}-\widehat{V}\widehat{U}|v\rangle +|d\rangle \langle h|\widehat{U}\widehat{V}+\widehat{V}\widehat{U}|v\rangle \right) . 
\end{align}
The fraction of the total intensity in port $x$ with polarization $y$ can then be evaluated as $P(x,y)=|\langle x,a|\Phi\rangle|^2$, resulting in 
%Since we are using a Dirac notation there is no notion of the absolute value of the intensity but we can calculate the fraction of the total intensity measured in a given port and polarization by the absolute value squared of the inner product of the $|\Phi\rangle$ and the measurement state as given by
\bea
   P(1,a) &=& %|\langle 1,a|\Phi\rangle|^2 
    \frac{1}{4}|\langle v| \widehat{U}\widehat{V}+\widehat{V}\widehat{U}|v\rangle|^2,
       \label{eq:Prob-1a}\\
   P(1,d) &=& %|\langle 1,d|\Phi\rangle|^2 
    \frac{1}{4}|\langle v|\widehat{U}\widehat{V}-\widehat{V}\widehat{U}|v\rangle|^2 ,
        \label{eq:Prob-1d}\\
    P(2,a) &=& %|\langle 2,a|\Phi\rangle|^2 
     \frac{1}{4}|\langle h| \widehat{U}\widehat{V}-\widehat{V}\widehat{U}|v\rangle|^2, 
        \label{eq:Prob-0a}\\
   P(2,d) &=& %|\langle 2,d|\Phi\rangle|^2 
    \frac{1}{4}|\langle h|\widehat{U}\widehat{V}+\widehat{V}\widehat{U}|v\rangle|^2 .
        \label{eq:Prob-0d}
\eea
%Here $P(x,y)$ is the fraction of the total intensity in port $x$ with polarization $y$. 
This is equivalent to the probability of detecting a photon at a given port with a given polarization, if we used a single photon set up. We note that in the experimental realization, these projections were performed by linear polarizers set to transmit diagonal and antidiagonal polarization positioned in front of CMOS cameras, as illustrated in Fig.~\ref{fig:experimental_setup}, but could be realized by a correctly oriented PBS followed by photo detectors to allow for a one shot measurement. 

In the following we outline how the geometric phase $\Omega/2$ related to the operation $\widehat{W}=\widehat{U}^\dagger \widehat{V}^\dagger \widehat{U}\widehat{V}$ can be obtained from the interferometer outputs. 
%Quantities (\ref{eq:Prob-1d})--(\ref{eq:Prob-0a}) are important because, as we show below, they are directly related to the geometric phase $\Omega/2$ for the series of transformations $\widehat{U}^\dagger \widehat{V}^\dagger \widehat{U}\widehat{V}$. 
Adding equations (\ref{eq:Prob-1d}) and (\ref{eq:Prob-0a}) and using the normalization condition $|h\rangle\langle h|+|v\rangle\langle v|=1$, we get
\bea \label{eq:p0a+p1d}
     P(1,d)+P(2,a) &=& \frac{1}{4}\langle v| (\widehat{U}\widehat{V}-\widehat{V}\widehat{U})^\dagger (\widehat{U}\widehat{V}-\widehat{V}\widehat{U})|v\rangle \nonumber \\
    %&=& \frac{1}{4}\langle v| \widehat{U}^\dagger \widehat{V}^\dagger \widehat{V}\widehat{U} + \widehat{V}^\dagger \widehat{U}^\dagger \widehat{U}\widehat{V} - \widehat{U}^\dagger \widehat{V}^\dagger \widehat{U}\widehat{V} - \widehat{V}^\dagger \widehat{U}^\dagger \widehat{V}\widehat{U}|v\rangle \nonumber \\
    %&=& \frac{1}{4} \left( 2 - \langle v| \widehat{U}^\dagger \widehat{V}^\dagger \widehat{U}\widehat{V} + \widehat{V}^\dagger \widehat{U}^\dagger \widehat{V}\widehat{U}|v\rangle \right) \nonumber \\
    &=& \frac{1}{2}- \frac{1}{2} \mathrm{Re}(\langle v| \widehat{U}^\dagger \widehat{V}^\dagger \widehat{U}\widehat{V}|v\rangle). \label{eq:Pcomb}
\eea
Now, let us take a state $|\psi\rangle$ such that $\widehat{U}\widehat{V}|\psi\rangle$ and $\widehat{V}\widehat{U}|\psi\rangle$ differ only by a phase factor, so that $\widehat{U}\widehat{V}|\psi\rangle = \e^{i\Omega/2}\widehat{V}\widehat{U}|\psi\rangle$, then
\begin{eqnarray}
 %\widehat{U}\widehat{V}|\psi\rangle &=& \e^{i\Omega/2}\widehat{V}\widehat{U}|\psi\rangle \\
    \widehat{W}|\psi\rangle=\widehat{U}^\dagger\widehat{V}^\dagger\widehat{U}\widehat{V}|\psi\rangle = \e^{i\Omega/2}|\psi\rangle\,.
\end{eqnarray}
This means that $|\psi\rangle$ is an eigenstate of $\widehat{W}$ with eigenvalue $\e^{i\Omega/2}$, and similarly the antipodal state $|\psi_\perp\rangle$ is the second eigenstate of $\widehat{W}$ with eigenvalue $\e^{-i\Omega/2}$. %The phase $\frac{\Omega}{2}$ is the geometric phase. 
Note that in general our actual experiment does not start at these eigenstates or follow the path $\widehat{W}=\widehat{U}^\dagger \widehat{V}^\dagger \widehat{U}\widehat{V}$. We find the geometric phase for $\widehat{W}$ using a fixed input state and following the actual closed path that generates the geometric phase is not required.
The states $|\psi\rangle$ and $|\psi\rangle_\perp$ form a basis in which we can write any polarization state. Specifically, let us write $|v\rangle = \alpha |\psi\rangle + \beta |\psi_\perp\rangle$ in order to relate $\Omega$ to the outputs of the Sagnac interferometer:
\begin{align}
     \langle v| \widehat{U}^\dagger \widehat{V}^\dagger \widehat{U}\widehat{V}|v\rangle & =  |\alpha|^2 e^{ i\Omega/2} + |\beta|^2 e^{ -i\Omega/2} \nonumber \\ & =\cos(\frac{\Omega}{2}) +i \sin(\frac{\Omega}{2})(|\alpha|^2  - |\beta|^2 )\,.
\end{align}
This means that equation (\ref{eq:Pcomb}) becomes 
\begin{equation}
 P(1,d)+P(2,a) = \frac{1}{2}-\frac{1}{2}\cos\left(\frac{\Omega}{2}\right)\,,
\end{equation}
which can be rearranged to give an expression for the geometric phase $\Omega/2$:
\begin{equation}
\frac{\Omega}{2} = \cos^{-1} \big(1-2P(2,a) -2P(1,d)\big) \label{eq:Omega}
\end{equation}
Note that this expression is independent of $\alpha$ and $\beta$, meaning that no prior knowledge of $|\psi\rangle$ and $|\psi_\perp\rangle$ -- and therefore of $\widehat{U}$ and $\widehat{V}$ -- is required to perform the measurement. 

At the same time, inserting Eq.~\ref{eq:Pcomb} into Eq.~\ref{eq:Omega} gives a description of the geometric phase in terms of the operators $\widehat{U}$ and $\widehat{V}$,
\begin{equation}
\frac{\Omega}{2} = \cos^{-1} \big(\mathrm{Re}(\langle v| \widehat{U}^\dagger \widehat{V}^\dagger \widehat{U}\widehat{V}|v\rangle)\big), 
\label{eq:OmegaJones}
\end{equation}
allowing us to evaluate the predict the geometric phase for waveplates with known Jones matrices. 

\begin{figure}
    \centering
\includegraphics[width=1\linewidth]{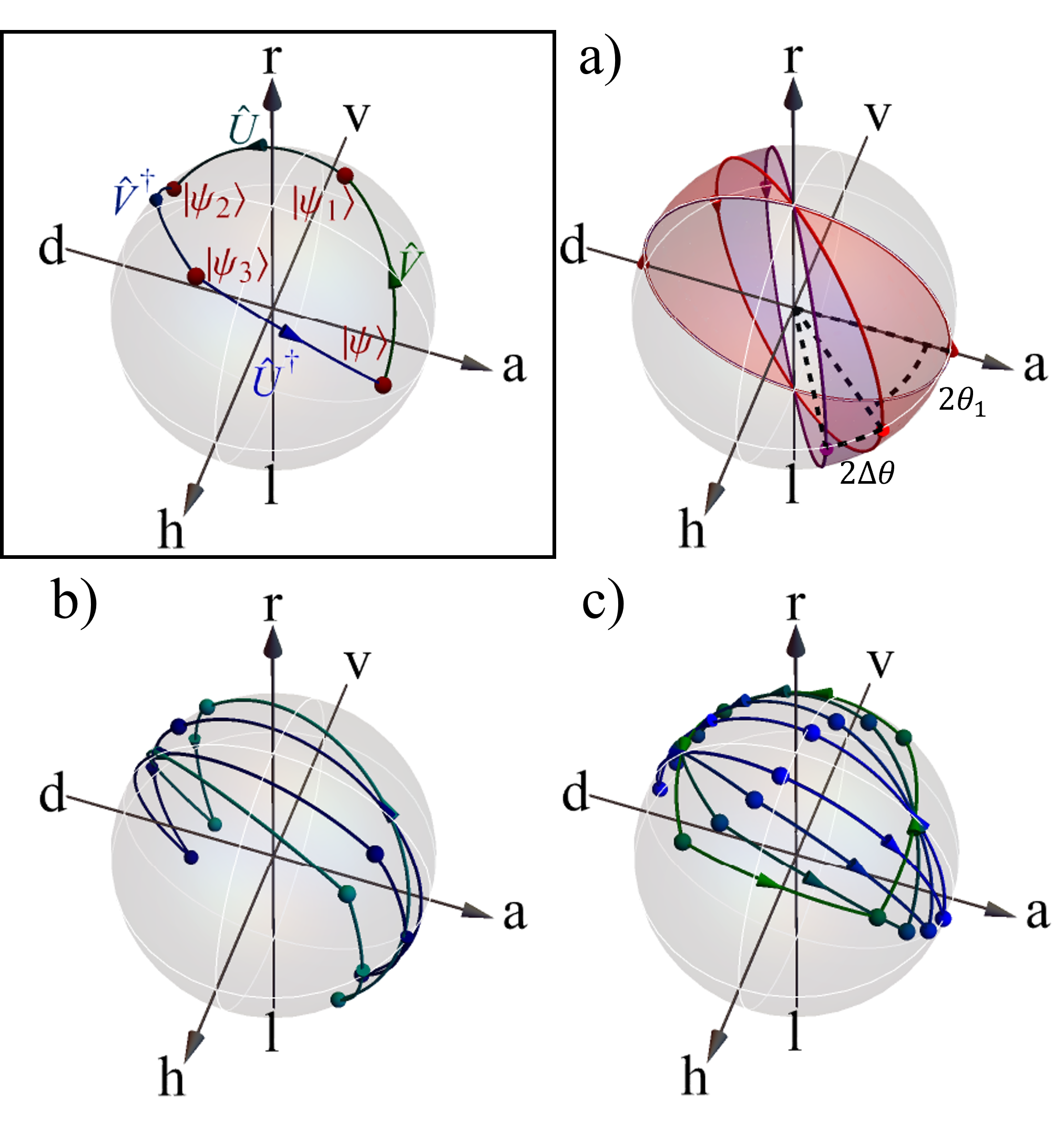}
    \caption{Illustrations of the Poincaré sphere representation of the geometric path corresponding to the sequential application of $\widehat{W}=\widehat{U}^\dagger \widehat{V}^\dagger \widehat{U}\widehat{V}$ enclosing a solid angle $\Omega$ of twice the geometric phase. %Each path starts and ends at a polarization eigenstate of $\widehat{W}$ for specific choices of $\widehat{U}$ and $\widehat{V}$. 
    \\ 
    Inset) Path for the case of $\widehat{U}=\mathrm{QWP}(0)$ and $\widehat{V}=\mathrm{QWP}(\pi/6)$, with notations as given in the main text.
    %. The path starts and ends at the eigenstate $|\psi\rangle=\widehat{U}^\dagger \widehat{V}^\dagger \widehat{U}\widehat{V}|\psi\rangle$, and proceeds via $|\psi_1\rangle=\widehat{V}|\psi\rangle$, $|\psi_2\rangle= \widehat{U}\widehat{V}|\psi\rangle$, and $|\psi_3\rangle= \widehat{V}^\dagger UV|\psi\rangle$. 
    a) Combination of a fixed HWP $\widehat{U}=\mathrm{HWP}(0)$ and a HWP $\widehat{V}=\mathrm{HWP}(\theta_1)$, with $\theta_1= \pi/9$ (shown in red) and $\theta_2=\pi/6$ (purple). The red shaded area shows $\Omega$ for $\theta_1$, and the purple shaded area indicates the linear increase in solid angle when $\theta$ is increased to $\theta_2$. 
    b) Combination of a fixed QWP $\widehat{U}=\mathrm{QWP}(0)$ and a HWP $\widehat{V}=\mathrm{HWP}(\theta_1)$ for $\theta_1=\pi/18$ (blue) and $\pi/9$ (green).
    c) Combination of two QWPs, one fixed $\widehat{U}=\mathrm{QWP}(0)$ and one with $\widehat{V}=\mathrm{QWP}(\theta_1)$, where $\theta_1 \in [\pi/18,\pi/9,3\pi/18, 2\pi/9]$ (shown in blue to green). }
    \label{fig:Poincare_Sphere}
\end{figure}

Figure~\ref{fig:Poincare_Sphere} illustrates the solid angle interpretation of geometric phase with several examples where the unitary matrices are chosen as half waveplates (HWPs) and quarter waveplates (QWPs) set at various angles. 
The path on the Poincaré sphere starts and ends at the eigenstate $|\psi\rangle=\widehat{W} |\psi\rangle$, and proceeds along geodesics via $|\psi_1\rangle=\widehat{V}|\psi\rangle$, $|\psi_2\rangle= \widehat{U}\widehat{V}|\psi\rangle$, and $|\psi_3\rangle= \widehat{V}^\dagger UV|\psi\rangle$. This is shown in the inset of Figure~\ref{fig:Poincare_Sphere} for a specific selection of waveplate settings. The enclosed solid angle $\Omega$ is twice the geometric phase associated with the operation $\widehat{U}^\dagger \widehat{V}^\dagger \widehat{U}\widehat{V}$, 
%The solid angle enclosed by a path starting at the eigenstate of the system then joining together the state after each operation by a geodesic on the Poincaré sphere is twice the geometric phase of the sequence of operations
\cite{Pancharatnam1956,BerryPhase1987,SolidAngleGeodesicVanDijk2010}. For any closed curve the solid angle can either be considered to enclose an inside area of less that $2\pi$ or enclose the rest of the sphere other than this area. We take the inside area in general as this is equivalent to choosing the positive phase in the eigenvalue problem and we do not measure the sign in our experiment only the absolute value of the solid angle $\Omega$.
A sequence involving HWPs is a special case as it can result in a transformation of a state to its antipodal point, with no unique geodesic connection. Mathematically this problem can be solved by  evaluating the limit as the HWP angle approaches the value of such angles, 
$\pi - \epsilon$, with $\epsilon$ set to 0.001 in our numerical calculations. 
%uses the limit as a waveplate approaches a HWP using a $\pi - 0.001$ rotation instead of a full $\pi$ rotation. This avoids issues with the Poincaré sphere picture for antipodal points. 
%Figure \ref{fig:Poincare_Sphere}a shows a solid angle formed from $\widehat{U}^\dagger \widehat{V}^\dagger \widehat{U}\widehat{V}$ when $\widehat{U}$ and $\widehat{V}$ are quarter waveplates (QWPs). 
Figure \ref{fig:Poincare_Sphere}a) shows that for a sequence of two HWPs, a particularly simple geometrical configuration emerges. In this case the eigenstate is fixed and the solid angle grows linearly with the angular difference between the fast axis of the two HWPs. The maximal value of $\Omega=2 \pi$ is reached when the path encloses half of the Poincaré sphere at $\Delta \theta = \pi/4$. 
This simple relationship breaks down for a sequence of two QWPs, as shown in Figure \ref{fig:Poincare_Sphere}b).  As the relative angle between the fast axis of QWPs increases, $\Omega$ changes non-linearly and moreover the eigenstates $|\psi\rangle$ change. This indicates the advantage of measuring the geometric phase using our ICO approach, as unlike conventional methods it does not require prior knowledge of the eigenstates. Finally, Figure \ref{fig:Poincare_Sphere}c) shows a sequence containing a QWP and a HWP. The enclosed area has an irregular shape that moves and changes nonlinearly with the angle between the waveplates in a more complicated version of the double QWP case. Our experiment allows the measurement corresponding to these solid angles for any pair of waveplates, or indeed other unitary operations, without requiring any prior knowledge.

It is instructive to contrast this scheme with established methods for measuring the geometric phase. Interferometric approaches based on Mach-Zehnder \cite{loredo2009measurement,garza2020geometric,kurzynowski2011geometric,machuca2026measurement, ortiz2014polarimetric,bhandari1988observation} or Michelson \cite{chyba1988measurement} geometries generally require precise control over the trajectory traced by the polarization state on the Poincaré sphere. In such measurements, a carefully prepared input state is subjected to a known sequence of unitary operations, and, since the total phase acquired is the sum of a geometric and a dynamical contribution, additional steps are typically needed to isolate the geometric part, for instance by enforcing a closed geodesic. In contrast, implementing an ICO protocol in a Sagnac interferometer as outlined here, measures the geometric phase directly, without any prior knowledge of the unitary operators. One potential caveat of our method is that it requires $\widehat{U}$ and $\widehat{V}$ to be unitary or losses to be path independent, whereas fringe-shift interferometers can also accommodate non-unitary filtering operations \cite{bhandari1988observation}. Unitary operations ensure that the two counter-propagating paths carry equal amplitude and the measured intensities isolate the geometric phase. We note however, that for specific applications it may be possible to compensate for differential losses by changing the weighting between the light inserted into the clockwise and anticlockwise Sagnac arms by careful design of the input state polarization.

\section{Experimental results and discussion}

We present experimental results demonstrating geometric phase measurement using the ICO protocol for three illustrative configurations of unitary transformations $\widehat{U}$ and $\widehat{V}$ realized by HWPs and QWPs at varying orientations, corresponding to the cases discussed in Figure \ref{fig:Poincare_Sphere}a), b) and c), for a sequence of two HWPs, two QWPs and a QWP and HWP, with varying angles of the fast-axis with respect to the vertical.
%: (a) HWP($\theta_1$) and HWP($\theta_2$), (b) HWP($\theta_1$) and QWP($\theta_2$), and (c) QWP($\theta_1$) and QWP($\theta_2$), where $\theta_{1,2}$ denotes the angle of the fast-axis of the waveplates with respect to the vertical.

As alignment of the Sagnac interferometer is critical, we optimized the overlap between clockwise and anticlockwise beams by monitoring the interference pattern on CMOS cameras. This was particularly important as rotation of the waveplates can introduce minute beam displacements which reduce the modulation of the interference pattern. 

The geometric phase in Eq.~(\ref{eq:Omega}) depends only on $P(1,d)$ and $P(2,a)$, i.e. the fraction of the total intensity measured in port 1 with diagonal polarization, and in port 2 with antidiagonal polarization. To ensure correct normalization and allowing for imperfect optical elements, we measure however the intensities in both ports for diagonal and antidiagonal polarization, resulting in 4 measures $I(1,a), I(1,d), I(2,a)$ and $I(2,d)$, from which we obtain the fractional intensities as    
$P(i,j)=I(i,j)/[I(1,a)+I(1,d)+I(2,a)+I(2,d)]$. 

For convenience we use the images from the CMOS cameras, corrected for background light to obtain the respective intensity values. For the QWP-QWP case, the intensity were additionally averaged over the relevant detector area, taken as a square with a side length of about a third of the beam waist. %$100\times100$ pixels. 

\begin{figure}
    \centering
\includegraphics[width=0.95\linewidth]{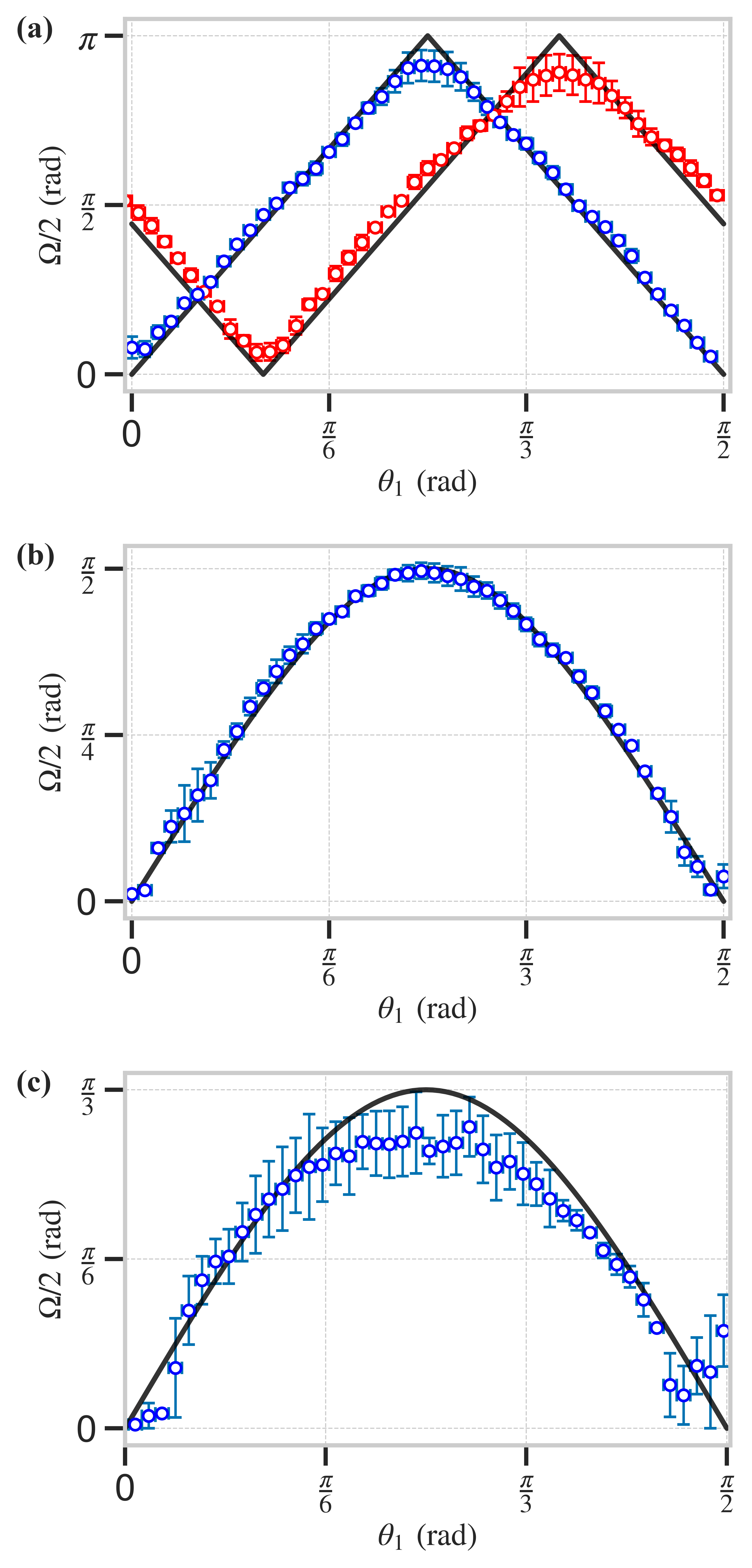}
    \caption{Measured and predicted geometric phases for different waveplate combinations. Data points derived from the ICO protocol are shown in blue or red, with vertical error bars indicating the standard error obtained for 3 measurement runs, and horizontal error bars derive from the alignment resolution. The corresponding predicted geometric phase for the given waveplates is shown as solid line.
    %The solid line represents the theoretical prediction,  while the blue circles with error bars correspond to the experimental measurements, given by $\bar{\Omega} \pm \sigma/\sqrt{N}$, where $\bar{\Omega}$ is the mean over the $N$ measurements and $\sigma/\sqrt{N}$ is the standard error. 
    Panel (a) shows the geometric phase arising for a configuration of  HWP($\theta$) and HWP($0$) (in blue) or  HWP($\pi/9$) (in red). Panel (b) shows HWP($\theta$) and QWP($0$), and panel (c) shows QWP$(\theta)$ and QWP($0$).}
    \label{fig:ico_results}
\end{figure}

Figure \ref{fig:ico_results} shows our experimental measurements of the solid angle together with the theoretical predictions for the geometric phase as a function of the fast-axis orientation of one of the waveplate performing the unitary operation $\widehat{V}$. 
%Each subfigure displays a different combination and configuration of waveplates. 
The data points shown in blue or red represent the average of $N = 3$ measurements 
% and the vertical error bars correspond to the standard error, while the horizontal error bars arise from the resolution of the measurement device. 
and the solid line represents the theoretical prediction for the geometric phase obtained from Eq.~\ref{eq:OmegaJones}, evaluated for the Jones vectors of the respective waveplates.
The solid angles for a waveplate of relative phase delay $\eta_1$ at an angle of 0 and a  waveplate of relative phase delay $\eta_2$ at an angle of $\theta$ are given by 
\begin{align}
\frac{\Omega}{2}=\arctan \sqrt{\left(\frac{4}{3+k_\eta+(1-k_\eta)\cos(4\theta)}\right)^2-1}, \label{eq:arctan}
\end{align}
where $k_\eta = \cos(\eta_1)+\cos(\eta_2)-\cos(\eta_1)\cos(\eta_2)$.

%This works well for QWP-QWP case and QWP-HWP case and works for HWP-HWP if you use ArcTan(tan(x)) = x but this does ignore the domain of Arctan. May need to consider a more general version for when the Solid angle is greater than Pi. This issue arises as the equation has come from an expression $e^{i\Omega/2} = \lambda$ for some unit magnitude eigenvalue therefore $\Omega/2 = -i\ln{(\lambda)}= -i(\ln{(|\lambda|}+i\arg{(\lambda)})=\arg{(\lambda)}$. This function has a range of $(-\pi,\pi)$. But when this is manipulated into a nicer form involving arctan this reduces to $(-\pi/2,\pi/2)$.

%The most general formula without this issue in its simplest form is given above $\Omega/2 = |\arg{(\lambda)}|$ where $\lambda = \frac{1}{4}(3+k_\eta + \cos (4\theta)(1-k_\eta)) \pm \frac{i}{4}\sqrt{16 -(3+k_\eta + \cos (4\theta)(1-k_\eta))^2 }$. Note the two eigenvalues given by the $\pm$ will give the two geometric phases differing by a sign so in order to get a positive $\Omega$ I take the absolute value which just selects one of the eigenstates. Also note that this form of the eigenstate makes it clear that $|\lambda| = 1$ as was required for $\ln{(|\lambda|})=0$

Figure \ref{fig:ico_results}(a) shows the results for a series of measurements of $\Omega$ for $\widehat{U}=\mathrm{HWP}(\theta)$ and $\widehat{V}=\mathrm{HWP}(0)$ (shown in blue) or $\widehat{V}=\mathrm{HWP}(\pi/9)$, where $\theta$ is rotated from $0$ to $\pi/2$. In agreement with our interpretation in terms of the Poincaré sphere shown in Figure \ref{fig:Poincare_Sphere}a), the geometric phase increases linearly until the solid angle encompasses a half sphere, and then decreases linearly. As for these configurations, the geometric phase depends solely on the differential orientation between the two HWPs, a different setting of $\widehat{V}$ simply results in a shift of the observed geometric phase. 
%Fig. \ref{fig:ico_results}(b) presents the results for the same set of wave-plates, but with $\theta_2$ fixed at $20^{\circ}$. It can be clearly observed that $\Omega$ is periodic in $\theta_1$ with period of $\pi/2$. 
Fig \ref{fig:ico_results}(b) and (c) show the results for a sequence of HWP and QWP, and for two QWPs, respectively. In both cases, the fast axes are initially aligned vertically. In both cases, the analytical evaluation of the Jones matrices predicts an arctan as given by Eq.~\ref{eq:arctan}, where
the maximum is reached when the fast and slow axes of the two waveplates are as far apart as possible.
%when the fast axis of both waveplates are perpendicular to each other. 

Our results are in good agreement with the theoretical predictions in all cases. The slight deviations from the expected values, observed primarily near the maxima, are mainly attributed to imperfections in the transmission properties of the PBS and the linear polarizer. The configuration containing two QWPs proved to be the noisiest. Firstly, the projection channels that feed directly into the measurement of $\Omega$, namely the terms $I(2,a)$ and $I(1,d)$, exhibit lower intensities, leading to a decrease in signal to noise ratio.
Secondly, retardation inhomogeneities across the beam profile of the waveplates cause the polarization state to vary spatially, so that regions near the beam edges are not fully canceled upon projection, resulting in residual interference fringes where ideally zero intensity is expected.

\section{Conclusion}

We have described and demonstrated experimentally that ICO protocols can be employed with Sagnac interferometers to measure the geometric phase associated with a sequence of unitary operations, without any prior knowledge of the individual operations.  
%The experimental setup successfully measures the geometric phase associated with $\widehat{U}^\dagger \widehat{V}^\dagger \widehat{U} \widehat{V}$ without any prior knowledge of $\widehat{U}$ or $\widehat{V}$ required. 
The experiments were performed with classical laser light, indicating that measurement advantages can persist in some instances beyond the quantum regime. 
This simple test case indicates that the Sagnac interferometer can give interesting and novel measurements that may be worth investigating further for more complicated transformations in particular where the order of operations is significant. ICO measurements may be advantageous for the identification of optical activity, for example the concentration of chiral fluids, especially for materials where exposure times have to be limited to avoid optical bleaching or denaturisation. A potential extension of this scheme involves testing with polarization structured light beams, which would allow the instant characterization of the full curves shown in Figure \ref{fig:ico_results} 
in a single measurement run.

\section*{Acknowledgments}

We are grateful to Tanmay Chandane who together with R.A.M built the initial setup that this work is based on. R.A.M. received funding via Fraunhofer CAP, award 322761-01. 
This study was financed in part by the Coordenação de Aperfeiçoamento de Pessoal de Nível Superior - Brasil (CAPES) - Finance Code 001. S. M. B. acknowledges support from the Royal Society, grant number RSRP/R/210005.  This work benefited from network activities through the INACT network, supported by the Engineering and Physical
Sciences Research Council (grant number EP/W026910/1).

\subsection*{Author Contributions}
\textbf{L.M.F.}: Data curation, Investigation, Formal analysis, Visualization, Writing – original draft.
\textbf{F.C.}: Formal analysis, Investigation, Methodology, Visualization, Software, Writing – original draft. 
\textbf{R.A.M.}: Data curation, Formal analysis. 
\textbf{R.M.dA.}: Supervision, Funding acquisition.
\textbf{S.C.}: Supervision, Conceptualization.
\textbf{S.M.B.}: Supervision, Conceptualization.
\textbf{S.F.-A.}: Conceptualization, Supervision, Writing – original draft, Funding acquisition, Resources.
All authors contributed to reviewing and editing the final manuscript.

\subsection*{Data Availability}
%The data that support the findings of this study are available from \cite{MarquesFagundes2026ICOcode} (to be replaced with the Enlighten Repository).
The data that support the findings of this study are available from the Enlighten Repository at \textbf{to be included} reference number \textbf{to be included}.

\bibliographystyle{apsrev4-2}
\bibliography{References}% Produces the bibliography via BibTeX.

\end{document}